# Generative models as parsimonious descriptions of sensorimotor loops

Manuel Baltieri and Christopher L. Buckley
EASY group – Sussex Neuroscience, Department of Informatics,
University of Sussex, Brighton, UK

m.baltieri@sussex.ac.uk        c.l.buckley@sussex.ac.uk
(+44 1273 876565)

*Abstract*

*The Bayesian brain hypothesis, predictive processing and variational free energy minimisation are typically used to describe perceptual processes based on accurate generative models of the world. However, generative models need not be veridical representations of the environment. We suggest that they can (and should) be used to describe sensorimotor relationships relevant for behaviour rather than precise accounts of the world.*

In the target article, Brette questions the use of the neural coding metaphor in the neurosciences. One of the main arguments is related to the criticism of approaches that overemphasise the role of perception as opposed to motor control for accounts of cognition. As suggested by Brette, while the sense-model-plan-act paradigm has long been criticised (Brooks, 1991), it still survives in modern approaches to neuroscience. He then extends this criticism to the notion of efficient coding and its most recent heir, predictive coding. Specifically, he argues that models that describe perception as a process of minimising redundancy (efficient coding) or prediction error (predictive coding) between incoming sensations and a generative model of sensory data are often used to ascribe the "goal" of minimising redundancy/error to an agent, in order to account for the rich repertoire of behaviours of living organisms. The shortcomings of this approach have previously been discussed, and resulted in the emergence of problems such as the dark-room paradox (Friston et al., 2012): why should agents show complex behaviour when, to minimise redundancy/prediction error, they could easily find the most predictable state in the world, e.g., a dark room?

Recently, the ideas of predictive coding/processing have been extended to include accounts of action, e.g., *active inference*. On this account, while perception is a process of changing predictions to better account for sensory data, action is a process of changing the world to better meet predictions. Normative behaviour then arises in this framework if an agent's generative model predicts rewarding states (Friston et al., 2012). Active inference moves the goal of cognitive agents from inferring properties of sensory data to acting in order to meet their goals. This extension of predictive coding is thus not in conflict with the ideas in the target article but, rather, directly supports them.

Furthermore, adaptive behaviour can emerge even if generative models are far divorced from a veridical representation of the environment. These *action-oriented* generative models, described in the context of *radical* predictive processing (Clark, 2015), can operate on the basis of approximate (e.g., linear) models of world dynamics (Baltieri & Buckley, 2019a) or simple sensorimotor couplings rather than objective representations of the world (Baltieri & Buckley, 2017). These ideas are derived from 4E (embodied, enactive, embedded and extended) theories of cognition which have long sought to address different issues of computationalism (Newen, et al., 2018), including the misuse of representational metaphors such as neural coding.

Thus, the idea of predictive processing/coding that the author describes, based on models that generate accurate representations of observed data and inherited from statistics and machine learning, is not the only game in town. Instead, we advocate for ideas of predictive coding that include generative models that are parsimonious descriptions of sensorimotor contingencies (Baltieri & Buckley, 2017). This may sound like an unnecessary stretch of the definition of a generative model (Bruineberg et al., 2018) but, we argue, is far from being just a semantic argument. The mathematical definition of these models is still entirely consistent with the more familiar notion of "generative" models and constitutes a valuable framework for the modelling of sensorimotor loops due to its strong and established relationships with (optimal) control theory (Todorov, 2009). In this context, "inference" can be best understood as a process of estimating actions necessary to attain future goals, *generating* expectations of desired states of affairs (rather than objective truths about the world) that are brought into existence by means of active behaviour.

In active inference, generative models further diverge in some fundamental ways from the more traditional ideal-observer-based forward models used in the context of motor neuroscience (Friston, 2011) and discussed in the target article. Forward models largely rely on a Kalman-like approach where all the variables affecting a system (parameters and inputs or causes, in a state-space formulation sense) and observations of said system (outputs) are available for an ideal observer to infer is latent states. On the other hand, generative models in active inference explicitly take the perspective of an agent into account. This includes discarding the idea that all causes affecting observations are known to an agent, suggesting instead the presence of approximate mechanisms to implement actions purely based on incoming sensations (not their estimates or "predictions") (Baltieri & Buckley, 2019b).

The target article further argues that theories based on Shannon communication theory (including efficient and predictive coding, but more in general all notions of neural coding) cannot in principle explain meaning in biological systems since this definition explicitly precludes a study of the semantics of information. We agree with Brette that notions of semantic information (with meaning for an agent, not necessarily for an experimenter) are largely neglected in neuroscience but his claim seems to overlook some of the efforts made to extend Shannon's work (Kolchinsky & Wolpert, 2018). To understand the implications of the value of information in biological systems, we suggest not to ignore frameworks based on Shannon information, but rather to further look into their connections to causal frameworks. In this light, extensions of predictive coding models such as active inference based on Bayesian networks could, in principle, describe biological agents as intervening in the world using motor actions to learn causal relationships between their actions and their sensations (Hohwy, 2013), i.e., the "subjective physics of the world" described by Brette.

The neural coding metaphor has long exhausted its appeal to explain cognition. This is because metaphors that attempt to describe living organisms as passively gathering and representing information from the environment cannot account for the complexity of the behaviour of biological systems. Similarly, we believe that a narrow interpretation of generative models as veridical world models is also a misdirection for cognitive science. Instead, we argue that the generative models at the heart of active inference, which are not to be seen as accurate maps of the world but as descriptions of valuable information and desires for an agent in the form of parsimonious actions/percepts relationships, will be a valuable tool for the study of cognitive systems.

**Acknowledgements**
MB would like to thank Martin Biehl and Simon McGregor for insightful discussions.